\documentclass[10pt,conference]{IEEEtran}
\usepackage{cite,graphicx,psfrag,amssymb,amsmath,subfigure,url}
\newtheorem{theorem}{Theorem}
\newtheorem{corollary}{Corollary}

\def\definedas{\triangleq}
\def\das{\triangleq}
\def\order{O}

\def\boldl{{\mbox{\boldmath $l$}}}
\def\boldsl{{\mbox{\scriptsize \boldmath $l$}}}
\def\Cost{L}
\def\p{p}
\def\q{\theta}

\def\sp{p}
\def\ll{(}
\def\lr{)}
\def\llex{\prec}
\def\period{{\mbox{.}}}
\def\comma{}
\def\L{{{\mathcal L}_n}}
\def\M{{\mathcal M}}

\def\X{{\mathcal X}}
\def\Z{{\mathbb Z}}
\def\lg{{\log_2}}

\begin{document}
\title{R\'{e}nyi to R\'{e}nyi --- Source Coding under Siege}
\author{\authorblockN{Michael B. Baer}
\authorblockA{Electronics for Imaging\\
303 Velocity Way\\
Foster City, California  94404  USA\\
Email: Michael.Baer@efi.com}}

\maketitle

\begin{abstract}
A novel lossless source coding paradigm applies to problems of
unreliable lossless channels with low bit rates, in which a vital
message needs to be transmitted prior to termination of
communications. This paradigm can be applied to Alfr\'{e}d R\'{e}nyi's
secondhand account of an ancient siege in which a spy was sent to
scout the enemy but was captured. After escaping, the spy returned to
his base in no condition to speak and unable to write. His commander
asked him questions that he could answer by nodding or shaking his
head, and the fortress was defended with this information. R\'{e}nyi
told this story with reference to prefix coding, but maximizing
probability of survival in the siege scenario is distinct from yet
related to the traditional source coding objective of minimizing
expected codeword length.  Rather than finding a code minimizing
expected codeword length $\sum_{i=1}^n p(i) l(i)$, the siege problem
involves maximizing $\sum_{i=1}^n p(i) \q^{l(i)}$ for a known $\q \in
(0,1)$.  When there are no restrictions on codewords, this problem can
be solved using a known generalization of Huffman coding.  The optimal
solution has coding bounds which are functions of R\'{e}nyi entropy;
in addition to known bounds, new bounds are derived here.  The
alphabetically constrained version of this problem has applications in
search trees and diagnostic testing.  A novel dynamic programming
algorithm --- based upon the oldest known algorithm for the
traditional alphabetic problem --- optimizes this problem in
$\order(n^3)$ time and $\order(n^2)$ space, whereas two novel
approximation algorithms can find a suboptimal solution faster: one in
linear time, the other in $\order(n \log n)$.  Coding bounds for the
alphabetic version of this problem are also presented.
\end{abstract}

\section{Introduction} 

Alfred R\'{e}nyi related an ancient scenario
in which the Romans held rebels under siege, rebels whose only hope
was the knowledge gathered by a mute, illiterate spy, one who could
only nod and shake his head \cite[pp.~13-14]{Reny}.  This apocryphal
tale --- based upon a historical siege --- is the premise behind the
Hungarian version of the spoken parlor game Twenty Questions.  A
modern parallel in the 21st century occurred when Russian forces
gained the knowledge needed to defeat hostage-takers by asking
hostages ``yes'' or ``no'' questions over mobile phones\cite{MSN,Tar}.

R\'{e}nyi presented this problem in narrative form in order to
motivate the relation between Shannon entropy and binary source
coding.  Note however that Twenty Questions, source coding, and the
siege scenario actually have three different objectives.  In Twenty
Questions, the goal is to be able to determine an item (or message) by
asking at most twenty questions.  In source coding, the goal is to
minimize the expected number of questions --- or, equivalently,
bits --- necessary to determine the message.  For the siege
scenario, the goal is survival, that is, assuming partial information
is not useful, the besieged would wish to maximize the probability
that the message is successfully transmitted within a certain window
of opportunity.  When this window closes and the siege ends, the
information becomes worthless.  An analogous situation occurs when a
wireless device is temporarily within range of a base station; one can
safely assume that the channel, when available, will transmit at the
lowest (constant) bitrate, and will be lost at a nondeterministic time
after its availability.

We consider this modified source coding problem and derive properties
of and algorithms for the optimization of the problem and 
variants thereof.  In Section~\ref{formal}, we formalize the problem
and find its solution in a generalization of the Huffman coding algorithm
previously used for a complementary problem.  Section~\ref{related}
concerns several extensions and variants of the problem.  In
particular, restricting the solution space to alphabetic codes is
considered in Section~\ref{alpha}, with a dynamic programming
algorithm presented for optimizing the alphabetic code, one that
extends to the related problem of search trees.  In
Section~\ref{simple}, we consider entropy bounds in the form of
R\'{e}nyi entropy for the unrestricted problem, leading to a new bound
and a related property involving the length of the shortest codeword
of an optimal code.  Entropy bounds for the alphabetic problem, along
with linear-time approximation algorithms, are derived in
Section~\ref{appx}.  Section~\ref{conclusion} concludes with related
work and a possible future direction.

\section{Formalizing the Problem} 
\label{formal}

A message is represented by symbol~$X$ drawn from the alphabet $\X
\das \{ 1, 2, \ldots, n \}$.  Symbol~$i$ has probability~$p(i)$,
defining probability mass function~$\p$, known to both sender and
receiver.  The source symbols are coded into binary codewords, each
bit of which is equivalent to an answer to a previously agreed-upon
``yes'' or ``no'' question; the meaning of each question (bit context)
is implied by the previous answers (bits), if any, in the current
codeword.  Each codeword~$c(i)$, corresponding to symbol~$i$, has
length~$l(i)$, defining overall length vector~$\boldl$ and overall
code~$C$.

Let~$\L$ be the set of allowable codeword length vectors, those that
satisfy the Kraft inequality, that is,
$$\L \definedas \left\{\boldl \in \Z_+^n \mbox{ such that }
\sum_{i=1}^n 2^{-l(i)} \leq 1\right\}\period$$ Furthermore, assume
that the duration of the window of opportunity is independent of the
communicated message and is memoryless.  Memorylessness implies that
the window duration is distributed exponentially.  Therefore,
quantizing time in terms of the number of bits $T$ that we can send
within our window, $$ P(T = t) = (1-\q)\q^t, ~ t = 0, 1, 2, \ldots $$
with known parameter $\q<1$.  We then wish to maximize the probability
of success, i.e., the probability that the message length does not
exceed the quantized window length:
\begin{eqnarray*}
P[l(X) \leq T] &=& \sum_{t=0}^\infty P(T=t) \cdot P[l(X) \leq t] \\
&=& \sum_{t=0}^\infty (1-\q)\q^t \cdot \sum_{i=1}^n p(i) 1_{l(i) \leq t} \\
&=& \sum_{i=1}^n p(i) \cdot (1-\q) \sum_{t=l(i)}^\infty \q^t \\
&=& \sum_{i=1}^n p(i) \q^{l(i)} \cdot (1-\q) \sum_{t=0}^\infty \q^t \\
&=& \sum_{i=1}^n p(i) \q^{l(i)}
\end{eqnarray*}
where $1_{l(i) \leq t}$ is $1$ if $l(i) \leq t$, $0$ otherwise.
The problem is thus the following optimization:
\begin{equation}
\max_{\boldsl \in \L} P[l(X) \leq T]  = \max_{\boldsl \in \L} \sum_{i=1}^n p(i) \q^{l(i)}
\label{one}
\end{equation}

To maximize this probability of success, we use a generalization of
Huffman coding developed independently by Hu {\it et
al.}~\cite[p.~254]{HKT}, Parker \cite[p.~485]{Park}, and Humblet
\cite[p.~25]{Humb0}, \cite[p.~231]{Humb2}.  The bottom-up algorithm of
Huffman coding starts out with $n$ weights of the form $w(i)=p(i)$ and
combines the two least probable symbols $x$ and $y$ into a two-node
subtree; for algorithmic reduction, this subtree of combined weights
is subsequently considered as one symbol with weight (combined
probability) $w(x) + w(y)$.  (We use the term ``weights'' because one
can turn a problem of rational probabilities into one of integer
weights for implementation.)  Reducing the problem to one with one fewer
item, the process continues recursively until all items are combined
into a single code tree.  The generalization of Huffman coding used to
maximize (\ref{one}) instead assigns the weight
$$\q \cdot (w(x) + w(y))$$ to the root node of the subtree of merged
items.  With this modified combining rule, the algorithm proceeds in a
similar manner as Huffman coding, yielding a code with optimal
probability of success.

\section{Related Problems}
\label{related}

Note that if we use this probability of success as a tie-breaker among
codes with minimal expected length --- those optimal under the
traditional measure of coding --- the solution is unique and
independent of the value of $\q$, a straightforward consequence of
\cite{Mark}.  We can obtain this optimal code by using the top-merge
variation of Huffman coding given in~\cite{Schw}; this variation views
combined items as ``smaller'' than individual items of the same
weight.  Similarly, for $\q$ sufficiently near $1$ --- i.e., if the
amount of information to be communicated is large compared to the size
of the window in question --- the optimal solution is identical to
this top-merge solution, a straightforward result analogous to that
noted in \cite[p.~222]{Knu1}.  Thus traditional Huffman coding should
be used if the window size is expected to be far larger than the
message size.

Observe also that, if we change the probability of $P(T=0)$ without
changing the ratios between the other probabilities, the problem's
solution code does not change, even though the probability of success
does.  There are still more criteria that are identically optimized, including,
if we have several independent messages serially transmitted, maximizing the
number of messages expected to be sent within a window.  Another
problem arises if we have a series of windows with independent
instances of the problem and want to minimize the expected numbers of
windows needed for success.  The maximization of probability minimizes
this number, which is the inverse of the probability of success in
each window:
\begin{equation*}
E[N_{\mbox{\scriptsize indep}}] = {\left(\sum_{i=1}^n p(i) \q^{l(i)}
\right)}^{-1}
\end{equation*}
Note that this is a risk-loving objective, in that we are more willing than
in standard coding to trade off having longer codewords for unlikely
items for having shorter codewords for likely items.

However, if the message to send is constant across all windows rather
than independent, the expected number of windows needed --- assuming
it is necessary to restart communication for each window --- is instead
$$E[N_{\mbox{\scriptsize const}}] = \sum_{i=1}^n p(i) \q^{-l(i)}
\period$$ 
This is a risk-averse objective, in that we are less willing to make
the aforementioned tradeoff than in standard coding.  These distinct
objective functions can be combined into one if we normalize, that is, if we seek
to minimize penalty function
\begin{equation}
\Cost_\q(\p,\boldl) \definedas \log_\q \sum_{i=1}^n
p(i)\q^{l(i)} 
\label{norm}
\end{equation}
for $\q>0$, where minimizing expected length is the limit case of $\q
\rightarrow 1$.  Campbell first noticed this in \cite{Camp1}.  Others
later found that the aforementioned generalized Huffman-like algorithm
optimizes this for all $\q>0$, though previously only $\q \geq 1$ had
any known application.

\section{Alphabetic Codes}
\label{alpha}

Under siege, assuming the absence of a predetermined code, using the
optimal Huffman-like code would likely be impractical, since
one would need to account not only for the time taken to answer a
question, but the time needed to ask it.  In this, and in applications
such as search trees and testing for faulty devices in a sequential
input-output system\cite{Yeu1} --- assuming the answer remains binary
--- each question should be of the form, ``Is the output greater than
$x$?'' where $x$ is one of the possible symbols, a symbol we call the
{\it splitting point} for the corresponding node.  This restriction is
equivalent to the constraint that $c(j) \llex c(k)$ whenever $j<k$,
where codewords $c(\cdot)$ are compared using lexicographical order.
The dynamic programming algorithm of Gilbert and Moore\cite{GiMo} can
be adapted to this restricted problem.

The key to the modified algorithm is to note that any optimal coding tree must
have all its subtrees optimal.  Since there are $n-1$ possible
splitting points, if we know all potential optimal subtrees for all possible
ranges, the splitting point can be found through sequential search of
the possible combinations.  The optimal tree is thus found inductively,
and this algorithm has $\order(n^3)$ time complexity and $\order(n^2)$
space complexity.  The dynamic programming algorithm involves finding
the maximum tree weight $W_{j,k}$ (and corresponding optimum tree) for
items $j$ through $k$ for each value of $k-j$ from $0$ to $n-1$,
computing inductively, starting with $W_{j,j} = w(j)$ ($=p(j)$), with
\begin{eqnarray*}
&W_{j,k} = \q \max_{s \in \{j,j+1,\ldots,k-1\}} [W_{j,s}+W_{s+1,k}]&
\end{eqnarray*}
for $j<k$.  Knuth showed how the traditional linear version of this
approach can be extended to general search trees\cite{Knu71}; for the
siege scenario, this is a straightforward generalization, which we
omit here for brevity.  For answers having unequal cost, algorithms
analogous to the linear-objective ones given in \cite{Itai,BaerCB} are
similarly formulated.

Another contribution of Knuth in \cite{Knu71} was to reduce
algorithmic complexity for the linear version using the fact that the
splitting point of an optimal tree must be between the splitting
points of the two (possible) optimal subtrees of size $n'-1$.  With
the siege problem, this property no longer holds; a counterexample to
this is $\q = 0.6$ with weights $(8, 1, 9, 6)$.

Similarly, for the linear problem\cite{HuTu}, as well as for $\q >
1$ and some nonexponential problems\cite{HKT}, there is a well-known
procedure --- the Hu-Tucker algorithm --- for finding an optimal
alphabetic solution in $\order(n \log n)$ time
and linear space.  The corresponding algorithm for $\q<1$ fails, however, this time
for $\q=0.6$ and weights
$(8, 1, 9, 6, 2)$.  Approximation algorithms presented in
Section~\ref{appx}, though, have similar or lesser complexity.

\section{Bounds on Optimal Codes}
\label{simple}

Returning to the general (nonalphabetic) case, it is often useful to
come up with bounds on the performance of the optimal code.  In this
section, we assume without loss of generality that $p(1) \geq p(2)
\geq \cdots \geq p(n)$.  Note that $\q \leq 0.5$ is a trivial case,
always solved by a unary code, $C_{\mbox{\scriptsize u}} \das (0, 10,
110, \ldots, 11{\cdots}10, 11{\cdots}11)$.  For nontrivial $\q >
0.5$, there is a relationship between the problem and R\'{e}nyi
entropy.

Campbell first proposed a decaying exponential utility
function for coding in \cite{Camp}.  He observed a simple upper bound
for (\ref{one}) with $\q>0.5$ in \cite{Camp} and alluded to a lower bound in
\cite{Camp0}.
These bounds are similar to the well-known Shannon entropy bounds for Huffman
coding (e.g., \cite[pp.~87-88]{CoTh}, \cite{Shan}).  In this case, however, the
bounds involve R\'{e}nyi's $\alpha$-entropy\cite{Ren2}, not Shannon's.
R\'{e}nyi entropy is
$$H_\alpha(\p) \definedas \frac{1}{1-{\alpha}}\lg \sum_{i=1}^n
 p(i)^{\alpha}\comma$$ where, in this case, $$\alpha \definedas \frac{1}{\lg
 2\q} = \frac{1}{1+\lg \q}\period$$

For nontrivial maximizations ($\q \in (0.5,1)$),
\begin{equation}
\q^{H_{\alpha}(\sp)+1} < \max_{\boldsl \in \L} P[l(X) \leq T] \leq
 \q^{H_{\alpha}(\sp)}. \label{bounds}
\end{equation}
We can rephrase this using the definition of 
$\Cost_\q(\p,\boldl)$ in (\ref{norm}) as
\begin{equation}
0 \leq 
\min_{\boldsl \in \L} \Cost_\q(\p,\boldl) - H_\alpha(\p) 
< 1,
\label{LH01}
\end{equation}
a similar result to the traditional coding bound\cite{Shan}.
Inequality~(\ref{LH01}) also holds for the minimization problem of $\q>1$.

As an example of these bounds, consider the probability distribution
implied by Benford's law\cite{Newc,Benf}:
\begin{equation}
p(i) = \log_{10}(i+1) - \log_{10}(i), ~ i = 1, 2, \ldots 9
\label{benf}
\end{equation}
At $\q=0.9$, for example, $H_\alpha(\p) \approx 2.822$, so the optimal
code will have between a $0.668$ and $0.743$ chance of success.
Running the algorithm, the optimal lengths are $\boldl = \ll 2, 2, 3,
3, 4, 4, 4, 5, 5 \lr$, resulting in a probability of success of
$0.739$.  

More sophisticated bounds on the optimal solution for the $\q>1$ case
were given in~\cite{BlMc}; these appear as solutions to related
problems rather than in closed form.  Closed-form bounds given in
\cite{Tane} are functions of entropy (of degree $\alpha$) and $p(1)$,
as in the linear case\cite{Gall,John,CGT,MoAb,CaDe1,Mans}.  These
bounds are flawed, however, in that they assume $p(1) \geq 0.4$ always implies
an optimal code exists with $l(1) = 1$.  A simple counterexample to this
assumption is $\p = (0.55, 0.15, 0.15, 0.15)$ with $\q = 2$, where
$l(i)=2$ for all~$i$.

However, when $\q < 1$, because the multiplication step of the
generalized Huffman-like coding algorithm provides for a strict reduction in
weight, $l(1)=1$ for any $p(1) \geq 0.4$.  Here we present better
conditions on $l(1)=1$ and show that they are tight, then derive
better entropy bounds from them.

\begin{theorem}
If $p(1) \geq 2\q(2\q+3)^{-1}$, then there is an optimal code for
$\p$ with $l(1)=1$.
\label{l1}
\end{theorem}

This is a generalization of \cite{John} and is only slightly more
complex to prove:

\begin{proof}
Recall that the generalized Huffman algorithm combines the items with
the smallest weights, $w'$ and $w''$, yielding a new item of weight $w
= \q(w'+w'')$, and this process is repeated on the new set of weights,
the tree thus constructed up from the leaves to the root.  Consider
the step at which item $1$ gets combined with other items; we wish to
prove that this is the last step.  At the beginning of this step the
(possibly merged) items left to combine are $\{1\}, S_2^k, S_3^k,
\ldots, S_k^k$, where we use $S_j^k$ to denote both a (possibly
merged) item of weight $w(S_j^k)$ and the set of (single) items
combined to make item $S_j^k$.  Since $\{1\}$ is combined in this
step, all but one $S_j^k$ has at least weight $p(1)$.  Recall too that
all weights $w(S_j^k)$ must be less than or equal to the sums of
probabilities $\sum_{i \in S_j^k} p(i)$.  Then
$$
\begin{array}{rclll}
\frac{2\q(k-1)}{2\q+3} &\leq& (k-1) p(1) && \\
&<& p(1) + \sum_{j=2}^k w(S_j^k) && \\
&\leq& p(1) + \sum_{j=2}^k \sum_{i \in S_j^k} p(i) && \\
&=& \sum_{i=1}^n p(i) &=& 1
\end{array}
$$ which, since $\q > 0.5$, means that $k < 5$.  Thus, because $n < 4$
is a trivial case, we can consider the steps in generalized Huffman
coding at and after which four items remain, one of which is item
$\{1\}$ and the others of which are $S_2^4$, $S_3^4$, and $S_4^4$.  We
show that, if $p(1) \geq 2/(2\q+3)$, these items are combined as shown
in Fig.~\ref{unary}.

\begin{figure}[ht]
     \centering
               \psfrag{X}{\mbox{\small $\X$}}
               \psfrag{p1}{\mbox{\small $\{1\}$}}
               \psfrag{S22}{\mbox{\small $S_2^2$}}
               \psfrag{S2}{\mbox{\small $S_2^4$}}
               \psfrag{W2}{\mbox{\scriptsize $\begin{array}{l} \mbox{If } |S_2^4|>1, \\ \quad w(S_2^4) = \\ \qquad \q [w(S'_2)+w(S''_2)]\end{array}$}}
               \psfrag{S2'}{\mbox{\small $S'_2$}}
               \psfrag{S2''}{\mbox{\small $S''_2$}}
               \psfrag{S3}{\mbox{\small $S_3^4$}}
               \psfrag{S4}{\mbox{\small $S_4^4$}}
               \psfrag{S3,4}{\mbox{\tiny $S_3^4 \cup S_4^4$}}
               \psfrag{W3,4}{\mbox{\scriptsize $\begin{array}{l} w(S_3^4 \cup S_4^4) = \\ \quad \q [w(S_3^4)+w(S_4^4)] \end{array}$}}
          \includegraphics{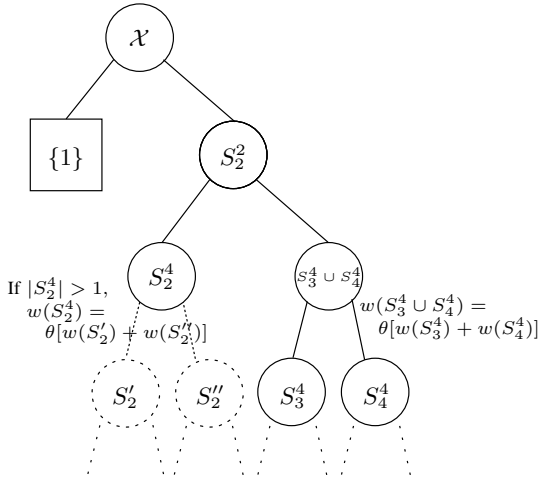}
     \caption{Tree in last steps of the generalized Huffman algorithm}
     \label{unary}
\end{figure}

We assume without loss of generality that weights $w(S_2^4)$,
$w(S_3^4)$, and $w(S_4^4)$ are in descending order.  From $w(S_2^4) +
w(S_3^4) + w(S_4^4) \leq \sum_{i=2}^n p(i) \leq 3/(2\q+3)$, $w(S_2^4)
\geq w(S_3^4)$, and $w(S_2^4) \geq w(S_4^4)$, it follows that
$w(S_3^4) + w(S_4^4) \leq 2/(2\q+3)$.  Consider set $S_2^4$.  If its
cardinality is $1$, then $w(S_2^4) \leq p(1)$, so the next step merges
the least two weighted items $S_3^4$ and $S_4^4$.  Since the merged
item has weight at most $2\q/(2\q+3)$, this item can then be combined
with $S_2^4$, then $\{1\}$, so that $l(1)=1$.  If $S_2^4$ is a merged
item, let us call the two items (sets) that merged to form it $S'_2$
and $S''_2$, indicated by the dashed nodes in Fig.~\ref{unary}.
Because these were combined prior to this step, $w(S'_2)+w(S''_2)
\leq w(S_3^4)+w(S_4^4)$, so $w(S_2^4) \leq \q [w(S_3^4)+w(S_4^4)] \leq
2\q/(2\q+3)$.  Thus $w(S_2^4)$, and by extension $w(S_3^4)$ and
$w(S_4^4)$, are at most $p(1)$.  So $S_3^4$ and $S_4^4$ can be
combined and this merged item can be combined with $S_2^4$, then $\{1\}$,
again resulting in $l(1)=1$.
\end{proof}

This can be shown to be tight by noting that
$$\p_\epsilon \definedas \left\ll\frac{2\q}{2\q+3}-3\epsilon,
\frac{1}{2\q+3}+\epsilon, \frac{1}{2\q+3}+\epsilon,
\frac{1}{2\q+3}+\epsilon\right\lr$$ has optimal length vector $\boldl =
\ll 2,2,2,2 \lr$ for any $\epsilon \in (0, (2\q-1)(8\q+12)^{-1})$.

Upper bounds derived from this, although rather complicated, are
improved. 

\begin{corollary}
For $l(1)=1$ (and thus for all $p(1) \geq 2\q(2\q+3)^{-1}$) and $\q<1$, the following holds:
$$\sum_{i=1}^n p(i) \q^{l(i)} > \q^2 \left[ 
\q^{\alpha H_\alpha(\p)} - p(1)^\alpha \right]^{\frac{1}{\alpha}} + \q p(1)$$
\end{corollary}

This is a straightforward consequence of Theorem~\ref{l1} and a proof is thus
omitted for space.  This upper bound is tight for $p(1) \geq 0.5$, as
$\p = (p(1), 1-p(1) +\epsilon, \epsilon)$ gets arbitrarily close for
small~$\epsilon$.

Let us apply this result to the Benford distribution in (\ref{benf})
for $\q=0.6$.  In this case, $H_\alpha(\p) \approx 2.260$ and $p(1) >
2\q(2\q+3)^{-1}$, so $l(1)=1$ and the probability of success is
between $0.251$ and $0.315=\q^{H_\alpha(\p)}$; the simpler 
(inferior) lower probability bound in (\ref{bounds}) is $0.189$.  The
optimal code is $\boldl = \ll 1,2,3,4,5,6,7,8,8 \lr$, which yields a
probability of success of $0.296$.

\section{Approximation Algorithms and Bounds for Alphabetic Codes}
\label{appx}

Returning again to alphabetic codes, if the dynamic programming
solution is too time- or space-consuming, an approximation algorithm
can be used.  A simple approximation algorithm involves adding one to
each of the lengths of an optimal nonalphabetic code; this yields
lengths corresponding to an alphabetic code, since $\sum_i 2^{-l(i)}
\leq 0.5$ is sufficient to have an alphabetic code\cite[p.~34]{AhWe},
\cite[p.~565]{Yeu1}.  Putting the lengths into (\ref{norm}),
$$\Cost_\q^{\mbox{\scriptsize huff}}(\p) \leq
\Cost_\q^{\mbox{\scriptsize alpha}}(\p) \leq
1+\Cost_\q^{\mbox{\scriptsize huff}}(\p)\comma$$ where
$\Cost_\q^{\mbox{\scriptsize huff}}(\p)$ is the cost of the optimal
code for the nonalphabetic problem.  Limits in terms of R\'{e}nyi
entropy follow from the previous section, and the following improved
approximation algorithm means that the right inequality is strict.

Approximation can be improved by utilizing techniques in
\cite{Yeu1} and \cite{Naka}.  The improved algorithm has two versions, one of
which is linear time, using the Shannon-like
$$l^\S(i) \definedas \left\lceil -\alpha \lg p(i) + \lg
\left(\sum_{j=1}^n p(j)^\alpha \right) \right\rceil \comma$$ and one
of which is $\order(n \log n)$ (or linear if sorting weights can be
done in linear time).

\begin{center}
{\bf Procedure for Finding a Near-Optimal Code}
\end{center}

\begin{enumerate}
\item Start with an optimal or near-optimal nonalphabetic code,
$\boldl^{\mbox{\scriptsize non}}$, such as the Shannon-like
$\boldl^{\mbox{\scriptsize non}}=\boldl^\S$ or the Huffman-like
$\boldl^{\mbox{\scriptsize non}}=\boldl^{\mbox{\scriptsize huff}}$.
\item Find the set of all minimal points, $\M$.  A minimal point is
any $i$ such that $1<i<n$, $l(i)<l(i-1)$, and $l(i)<l(i+1)$.
Additionally, if $l(i-1)>l(i)=l(i+1)=\cdots = l(i+k) < l(i+k+1)$,
then, of these, only $j \in [i, i+k]$ minimizing $w(j)$ (or $p(j)$) is
a minimal point.
\item Assign a preliminary alphabetic code with lengths
$l^{\mbox{\scriptsize pre}}=l^{\mbox{\scriptsize non}}+1$ for all
minimal points, and $l^{\mbox{\scriptsize pre}}=l^{\mbox{\scriptsize
non}}$ for all other items.  This corresponds to an alphabetic code
$C^{\mbox{\scriptsize pre}}$.  Note that such an alphabetic code is
easy to construct; the first codeword is $l(1)$ zeros, and each
additional codeword $c(i)$ is obtained by either truncating $c(i-1)$
to $l(i)$ digits and adding $1$ to the binary representation (if
$l(i)\leq l(i-1)$) or by adding $1$ to the binary representation of
$c(i-1)$ and appending $l(i)-l(i-1)$ zeros (if $l(i) > l(i-1)$).
\item Go through the code tree (with, e.g., a depth-first search), and
replace any node having only one child with its grandchild or
grandchildren.  At the end of this process, an alphabetic code with
$\sum_{i=1}^n 2^{-l(i)}=1$ is obtained.
\end{enumerate} 

This hybrid of the approaches of Nakatsu\cite{Naka} and
Yeung\cite{Yeu1} can be easily applied to all $\q>0$, including the
linear limit case, for which it is an improved approximation technique
when $\boldl^{\mbox{\scriptsize non}}=\boldl^{\mbox{\scriptsize
huff}}$.

\section{Related Work, Extensions, and Conclusion}
\label{conclusion}

The algorithms presented here will not work if $n=\infty$, although
methods are known of finding codes for geometric and lighter
distributions\cite{Baer07} and existence results are known for all
finite-(R\'{e}nyi) entropy distributions\cite{Baer06}.  Also, although
presented here in binary form for simplicity's sake, nonalphabetic
results readily extend to $D$-ary codes \cite{Camp0, Camp, Humb2}.
The alphabetic algorithm extends in a manner akin to that shown for the
extension of the Gilbert and Moore algorithm
in~\cite[pp.~15-16]{Itai}.
Further upper bounds on optimal $\Cost_\q(\p, \boldl)$ are elusive,
but should be quite similar to those for the linear case, at least for
$\q < 1$, since the distributions approaching or achieving these
bounds should be of bounded cardinality almost everywhere.

In conclusion, when R\'{e}nyi's siege scenario is formalized, problem
solutions involve Huffman coding, dynamic programming, and,
appropriately, R\'{e}nyi entropy.

\section*{Acknowledgments}

The author would like to thank Thomas M. Cover, T. C. Hu, and J. David
Morgenthaler for discussions and encouragement on this topic, as well
as the two anonymous reviewers for suggestions on presentation.  The
author was partially supported, while at Stanford University in the
initial phase of this research, by the National Science Foundation
(NSF) under Grant CCR-9973134 and the Multidisciplinary University
Research Initiative (MURI) under Grant DAAD-19-99-1-0215.

\end{document}